\documentclass[aps,pre,superscriptaddress,twocolumn,showpacs]{revtex4}
\usepackage{amsmath,amssymb,amsfonts}
\usepackage{bbm}
\usepackage{graphicx}
\newcommand\ImgI{\mathrm{i}}

\begin{document}

\title{Universal spectral statistics of Andreev billiards:
  semiclassical approach}

\author{Sven Gnutzmann}
\email{gnutz@physik.fu-berlin.de}
\affiliation{Institut f\"ur Theoretische Physik, Freie Universit\"at
  Berlin, Arnimallee 14, 14195 Berlin, Germany} \author{Burkhard Seif}
\email{bseif@thp.uni-koeln.de} 
\affiliation{Institut f\"ur Theoretische Physik, Universit\"at zu
  K\"oln, Z\"ulpicher Str.\ 77, 50937 K\"oln} \author{Felix von Oppen}
\email{vonoppen@physik.fu-berlin.de} 
\affiliation{Institut f\"ur Theoretische Physik, Freie Universit\"at
  Berlin, Arnimallee 14, 14195 Berlin, Germany} \author{Martin R.
  Zirnbauer}
\email{zirn@thp.uni-koeln.de} 
\affiliation{Institut f\"ur Theoretische Physik, Universit\"at zu
  K\"oln, Z\"ulpicher Str.\ 77, 50937 K\"oln}
  
\begin{abstract}   
  The symmetry classification of complex quantum systems has recently
  been extended beyond the Wigner-Dyson classes.  Several of the
  novel symmetry classes can be discussed naturally in the context of 
  superconducting-normal hybrid systems such as Andreev billiards and 
  graphs. In this paper, we give a semiclassical interpretation of 
  their universal spectral form factors in the ergodic limit.
\end{abstract}      


\maketitle

\section{Introduction}

Based on early work of Wigner \cite{Wigner}, Dyson \cite{Dyson}
proposed a classification of complex quantum systems according to
their behavior under time reversal and spin rotations.  The ergodic
limits of the proposed symmetry classes are described by the Gaussian
orthogonal, unitary, and symplectic ensembles (GOE, GUE, and GSE) of
random-matrix theory.  These were initially motivated by atomic
nuclei and have since been applied successfully to a large variety of
systems, most notably chaotic and disordered quantum systems
\cite{Guhr}.  More recently, an additional seven symmetry classes have
been identified \cite{Zirnbauer}, which are naturally realized in part
by Dirac fermions in random gauge fields (chiral classes)
\cite{chiral} and in part by quasiparticles in disordered mesoscopic
superconductors \cite{asz} or super\-con\-duc\-ting-normalconducting
(SN) hybrid systems \cite{Altland}. The common new feature of the
new symmetry classes is a mirror symmetry in the spectrum: if $E$
is in the spectrum, so is $-E$.
The corresponding Gaussian 
random-matrix ensembles differ from
the Wigner-Dyson ensembles in so far as their spectral statistics, while
still universal, is no longer stationary under shifts of the energy
due to additional discrete symmetries.

Much insight into the range of validity of the Wigner-Dyson
random-matrix ensembles has been gained from the semiclassical
approach to the spectral statistics of chaotic quantum systems, based
on Gutzwiller's trace formula \cite{Gutzwiller}. 
In a seminal paper \cite{Berry}, Berry
gave a semiclassical derivation of the spectral form factor of chaotic
quantum systems for the Wigner-Dyson ensembles, partially reproducing
the results of random-matrix theory
and clarifying its limitations. In this paper we provide such a
semiclassical interpretation, based on Andreev systems, of a
generalized form factor for the Gaussian random-matrix ensembles
associated
with the new symmetry classes termed $C$ and $C$I (the pertinent
Gaussian ensembles will be referred to as $C$-GE and $C$I-GE).

There have been several attempts to apply semiclassical theory to SN
hybrid systems \cite{Melsen,Taras,Richter}. Melsen {\it et al.}
\cite{Melsen} pointed out that the gap induced by the proximity effect
in a billiard coupled to a superconducting lead (Andreev billiard --
cf.\ Fig.\ 1b) is sensitive to whether the classical dynamics of the
(normal) billiard is integrable or chaotic. These authors showed that
the proximity-induced hard gap in the chaotic case is \emph{not}\,
fully reproduced by semiclassical theory, the reasons for which have
been discussed further in Ref.\ \cite{Taras}. 
All these systems have in common that due to the presence of the 
superconductor the combined electron-hole dynamics is no longer chaotic
even if the corresponding normal (nonsuperconducting) billiard is chaotic
\cite{Andreevbilliard}.
By contrast, we focus
here on a semiclassical approach to SN hybrid systems where the combined
electron-hole dynamics remains chaotic even in the presence of
the superconductor. Such systems exhibit
the \emph{universal} spectral statistics of the Gaussian random-matrix 
ensembles for the new symmetry classes.  
We identify the class of periodic orbits contributing to universal
features in the density of states near the Fermi energy
and show that in this case, semiclassics reproduces the
spectral statistics predicted by random-matrix theory.

\section{Universal spectral statistics}
\label{sec:rmtsummary}

We briefly summarize the pertinent random-matrix results for the Gaussian
ensembles corresponding to the new
symmetry classes. For the Wigner-Dyson ensembles (GUE, GOE, GSE), 
the average density of
states is nonuniversal and random-matrix theory
makes
universal predictions only about spectral fluctuations 
in the ergodic limit such as the
correlation function 
\begin{equation}
  C(\epsilon) = \langle \delta \rho(E) \delta
  \rho(E + \epsilon)\rangle
\end{equation}
of the deviations $\delta\rho(E)$ of the
density of states $\rho(E)$ from its mean value $\langle \rho(E)
\rangle$.  A central quantity is the spectral form factor 
\begin{equation}
  K_{\text{WD}}(t) = \frac{1}{\langle\rho\rangle} 
  \int_{-\infty}^\infty d\epsilon\,
  e^{-\ImgI \epsilon t/ \hbar} C(\epsilon).
\end{equation}
The ergodic limit of the new
symmetry classes differs from the Wigner-Dyson case by the fact that
even the average density of states has universal features close to the
Fermi energy $\mu$. Thus, in this case, we define a generalized
spectral form
factor by the Fourier transform of the expectation value
of the (oscillating part of the) density of states:
\begin{equation}
  K(t) = 2\, \int_{-\infty}^{\infty} dE \langle \delta \rho(E)\rangle
  e^{-\ImgI E t/\hbar},
\end{equation} where $E$ is the energy (measured relative to
$\mu$). 
For the ensemble $C$-GE (class $C$ is 
invariant under spin rotations, while time
reversal is broken), this form factor is \cite{Altland}
\begin{equation}
  K^{C}(t) = -\theta (1-\frac{|t|}{t_H}).
\label{rmt-result}
\end{equation}
Here, $t_H = 2\pi\hbar \rho_{\mathrm{av}}$ is the Heisenberg time defined in
terms of the mean density of states  $\rho_{\mathrm{av}}$ sufficiently far from the Fermi
energy (the oscillating part of the density of states is 
defined as $\delta\rho(E)=\rho(E)-\rho_{\text{av}}$). 
Semiclassically $\rho_{\mathrm{av}}$ corresponds to
Weyl's law. For the ensemble $C$I-GE (class $C$I differs from $C$ by invariance
under time-reversal), the short-time expansion is \cite{Altland}
\begin{equation}
  K^{C\mathrm{I}} (t) = -1+\frac{|t|}{2t_H}+\mathcal{O}(|t|^2).
\end{equation}
Wigner-Dyson statistics can be applied even to a single chaotic
system 
by exploiting a spectral average. 
By contrast, the new
symmetry classes require an ensemble average since
they have universal features 
in the vicinity of special energies (Fermi energy $\mu$).
For billiards one may average 
over shapes.

Before entering into the semiclassical analysis for the new symmetry
classes, we briefly review the semiclassical derivation of the usual
spectral form factor of the GUE. There one starts from the Gutzwiller
trace formula, that relates the oscillatory contribution
$\delta\rho(E)$ to the density of states to a sum over periodic orbits
$p$, 
\begin{equation}
  \delta\rho(E) = \frac{1}{\pi\hbar} \mathrm{Re} \sum_p t_p A_p
  e^{\ImgI S_p/\hbar}.
\end{equation}
Here, $S_p$ denotes the classical action of the
orbit, $A_p$ denotes its stability amplitude, and $t_p$ is the
primitive orbit traversal time. The explicit factor $t_p$ arises
because the traversal of the periodic orbit can start anywhere along
the orbit.  Inserting this expression into the definition of the
spectral form factor, and employing the diagonal approximation, one
finds 
\begin{equation}
  K_{\mathrm{WD},\mathrm{diag}}(t) = \sum_p \frac{t_p^2}{t_H} |A_p|^2
\delta(t-t_p).
\end{equation}
Finally averaging over some time interval $\Delta t$
and using the Hannay--Ozorio-de-Almeida sum rule \cite{sumrule}
$\sum_p^{t_p \in [t,t+\Delta t]} |A_p|^2 = \Delta t/t$ one obtains the
result 
\begin{equation}
  K_{\mathrm{WD},\mathrm{diag}}(t)= \frac{t}{t_H}
\end{equation} 
valid for $t_0\ll t\ll t_H$, where $t_0$ is the period 
of the shortest periodic 
orbit. This result agrees with the short-time behavior of the spectral form
factor predicted by the GUE.

\section{Semiclassical approach to magnetic Andreev billiards}
\label{sec:billiards}

We now turn to Andreev billiards -- the central theme of this paper.
The novel element in SN hybrid systems is Andreev reflection
converting electrons into holes (and vice versa) at the interface to
the superconductor (see Fig.\ 1a).  In this process, the incoming
electron (hole) acquires a phase $-\ImgI e^{-\ImgI\alpha}$ ($-\ImgI
e^{\ImgI\alpha}$), where $\alpha$ is the phase of the superconducting
order parameter $\Delta$ \cite{gauge}. In the absence of a magnetic field,
electrons (holes) sufficiently close to the Fermi energy 
($E\ll |\Delta| \ll \mu$) are reflected
as holes (electrons) which then retrace the electron (hole) trajectory
backwards (retroflection). 
In chaotic billiards, essentially all
trajectories eventually hit any given part of the boundary.
Thus, if the billiard is then coupled to a superconductor
any quasiparticle eventually hits the superconducting interface, 
leading to a periodic orbit bouncing back and forth between two points
on the superconducting interface.  It follows that a conventional chaotic
billiard (without magnetic field) that is coupled to a superconductor
has a combined electron-hole dynamics that is no longer chaotic.
Instead, the resulting trajectories are all periodic, leading to
nonuniversal behavior such as the proximity-induced hard gap
\cite{Melsen,Taras} for time-reversal invariant systems.

\begin{figure}
  \begin{center}
    \includegraphics[width=0.47\linewidth]{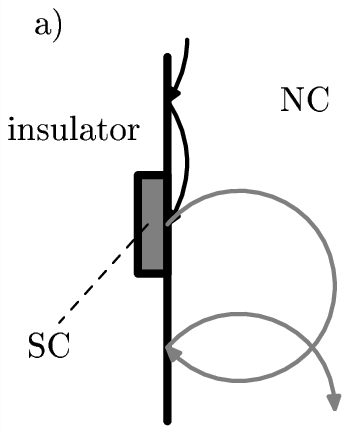}\;
    \includegraphics[width=0.47\linewidth]{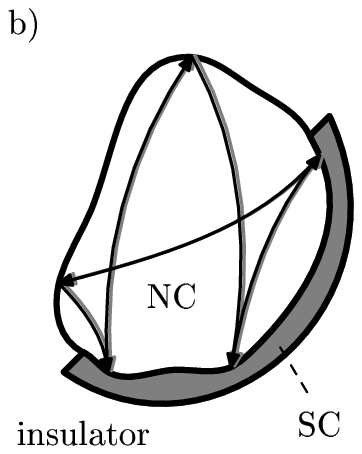}
    \caption{
      (a) Andreev scattering with a perpendicular magnetic field.  (b)
      Andreev billiard -- a part of the boundary is connected to a
      superconductor where Andreev reflection (retroflection)
      takes place. An example
      of a self-dual orbit is shown -- the SN interface is hit three
      times by an incoming electron and three times by an incoming
      hole. Each part of the trajectory is traversed twice in the same
      direction -- once as an electron and once as a hole. The singly
      traversed orbit without electron-hole labels is a periodic orbit
      of the virtual billiard and hits the SN interface three
      times.}
\end{center}
\end{figure}

One expects to recover universal spectral statistics \emph{only} if
the \emph{combined} electron-hole dynamics is chaotic and periodic
orbits are isolated as in conventional chaotic (hyperbolic) systems.
In Andreev billiards this occurs naturally when time-reversal symmetry
is broken by a perpendicular magnetic field (symmetry class $C$). 
In this case
the retroflected hole (electron) does not retrace the trajectory of
the incoming electron (hole), as  electron and hole trajectories
are curved in the same direction (cf.\ Fig.\ 1a).  This allows one to
express the density of states semiclassically
by a Gutzwiller-type trace formula as a
sum over the isolated periodic orbits of the \emph{combined}
electron-hole dynamics of the Andreev billiard,
\begin{equation}
  \delta\rho(E)= \frac{1}{\pi\hbar} \mathrm{Re} \sum_p t_p A_p e^{\ImgI
    S_p(E)/\hbar+\ImgI\chi}.  
\end{equation}
The orbit amplitudes $A_p$ are products of electron and hole
contributions, 
\begin{equation}
  A_p = A_p^{(e)} A_p^{(h)},
\end{equation}
while the orbit actions
are sums of electron and hole actions, 
\begin{equation}
  S_p(E) = S_p^{(e)}(E) +
  S_p^{(h)}(E).
\end{equation}  
The factor $t_p$ again reflects the arbitrary starting
point of the orbit and $\chi$ denotes the accumulated Andreev phases.

Coherent contributions to the form factor can be expected from the
periodic orbits that retrace the same trajectory in the same direction
with the roles of electrons and holes interchanged. Such
\emph{self-dual} orbits are invariant under electron-hole conjugation
and the dynamical contributions to their action largely cancel due to
the relation $S_p^{(e)}(E) = -S_p^{(h)}(-E)$, so that $S_p(E)\simeq
E t_p$.  Moreover, the amplitudes of electron and hole are just complex
conjugates of one another, giving $A_p=|A_p^{(e)}|^2$. The accumulated
Andreev phase is $(-\ImgI)^{2s}=-1$ with $s$ an odd integer.  Keeping
only the self-dual periodic orbits -- the \emph{self-dual
  approximation} -- we find 
\begin{equation}
  \delta\rho(E)_{\mathrm{sd}}=
  -\frac{1}{\pi\hbar} \mathrm{Re}\sum_p t_p |A_p^{(e)}|^2 e^{\ImgI E
    t_p/\hbar}.
\end{equation}
For the generalized form factor, this leads to
\begin{equation}
  K(t)_{\mathrm{sd}}= - 2
  \sum_{p:\text{sd}} t_p |A_p^{(e)}|^2 \delta(t-t_p).
  \label{sd}
\end{equation}
This expression reveals the similarity to the diagonal approximation
for the Wigner-Dyson form factors.  However, only \emph{one} factor
$t_p$ arises. 

The Hannay--Ozorio-de-Almeida sum rule does not apply directly 
to the sum over self-dual
orbits (being a sum over amplitudes of a subclass of periodic orbits). 
To deal with this difficulty, we introduce a \emph{virtual
  billiard} with the same dynamics as the Andreev billiard except that
there is \emph{no} particle-hole conversion at the SN interface.
Thus, the virtual billiard is an ordinary chaotic billiard with
unusual reflection conditions at the SN interface (retroflections).
Primitive periodic orbits of the virtual billiard involve either even
or odd numbers of retroflections. Reintroducing electron-hole
conversion, one observes that even orbits lead to non-self-dual
periodic orbits in the Andreev billiard. By contrast, twofold traversals of
odd orbits are periodic and self dual in the Andreev billiard as
the roles of electron and hole are interchanged in the second
traversal (see Fig.\ 1b). We can now interpret the sum 
over self-dual orbits in Eq.\
(\ref{sd}) as a sum over odd orbits of the virtual billiard. Since on
average half of its orbits are odd, the Hannay--Ozorio-de-Almeida
sum rule for the virtual
billiard gives 
\begin{equation}
  \sum_{p:\text{sd}}^{t_p \in [t,t+\Delta t]} |A^{(e)}_p|^2 = \frac{\Delta
    t}{2t}.
\end{equation}
Thus
\begin{equation}
  K(t)_{\mathrm{sd}}=  -1
\end{equation}
in agreement with the random-matrix result 
predicted by $C$-GE (\ref{rmt-result}) for short times.
The self-dual approximation is expected to hold for
$t_{0},t_{\mathrm{A}}\ll t\ll t_H$
where $t_0$ is the traversal time
of the shortest periodic orbit and
$t_{\mathrm{A}}$ the Andreev time (typical
time until electron-hole conversion takes place).

\section{Spectral statistics for Andreev graphs}
\label{sec:graphs}

The semiclassical calculation of the form factor becomes particularly
transparent and explicit for quantum graphs which were recently
introduced \cite{ksmila} as simple quantum chaotic systems.
Introducing Andreev reflection as a new ingredient, we show
\emph{semiclassically} that the form factor of the resulting
\emph{Andreev graph} takes on the universal result.  A quantum graph
consists of vertices connected by bonds. A particle (electron/hole)
propagates freely on a bond and is scattered at a vertex according to
a prescribed scattering matrix.  For definiteness, we discuss star
graphs with $N$ bonds of equal length $L$. These have one
\emph{central} vertex and $N$ \emph{peripheral} vertices.  Each bond
connects the central vertex to one peripheral vertex (cf.\ Fig.\ 2).

\begin{figure}
  \begin{center}
    \includegraphics[width=0.6\linewidth]{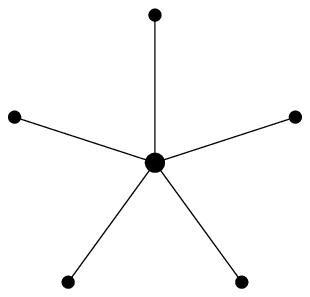}
    \caption{
      Star graph with five peripheral vertices connected
      to superconductors.}
\label{fig:Andreevbill}
\end{center}
\end{figure}

Andreev (star) graphs are obtained by introducing (complete)
electron-hole conversions at the peripheral vertices, while the central
vertex preserves the particle type.  The quantization condition is
\begin{equation}
  \det \left( \mathcal{S}(k)- \mathbbm{1} \right)=0, 
\end{equation}
with the unitary
$N \times N$ matrix 
\begin{equation}
  \mathcal{S}(k)= S_C \mathcal{L} D_{-} \mathcal{L}
  S_C^* \mathcal{L} D_{+} \mathcal{L}.
\end{equation}  
Here $S_C$ ($S_C^*$) is the
central scattering matrix for an electron (hole).  
For definiteness, we choose \cite{Tanner}
\begin{equation}
  S_{C,kl}
  = \frac{1}{\sqrt{N}}e^{2\pi\ImgI kl/N},
\end{equation} 
where $S_C$ by itself does not
break time-reversal symmetry.  The matrix 
\begin{equation}
  \mathcal{L} = e^{\ImgI k
    L}\mathbbm{1}
\end{equation}
contains the phases accumulated when the
quasiparticle propagates along the bonds ($k$ is the wave number
measured from the Fermi wave number).  Finally, 
\begin{equation}
  D_{\pm} = -\ImgI\,
  \mathrm{diag}(e^{\mp\ImgI \alpha_i})
\end{equation}
contains the \emph{Andreev phases}
accumulated at the vertices, where $\alpha_i$ denotes the
order-parameter phase at peripheral vertex $i$.  Time-reversal
symmetry is obeyed if all Andreev phases are either $\alpha_i=0$ or
$\alpha_i=\pi$, but is broken otherwise. Accordingly, we build
ensembles corresponding to the symmetry classes $C$ (uncorrelated
Andreev phases $\alpha_i$ with uniform distributions in the interval
$\left[0,2\pi\right)$) and $C$I (uncorrelated Andreev phases taking
values $\alpha_i=0$ or $\alpha_i = \pi$ with equal probability).
Numerically computed ensemble averages are in excellent agreement with
random-matrix results from $C$-GE and $C$I-GE, as shown in Fig.\ 3.

\begin{figure}
  \begin{center}
    \includegraphics[width=0.95\linewidth]{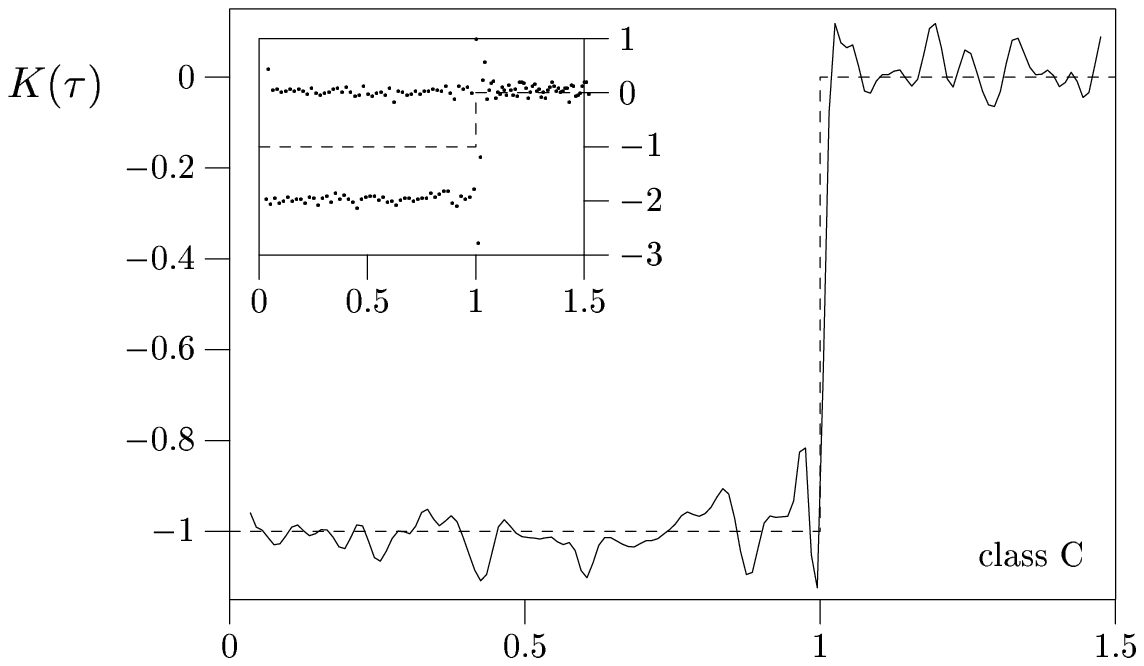}
    \includegraphics[width=0.95\linewidth]{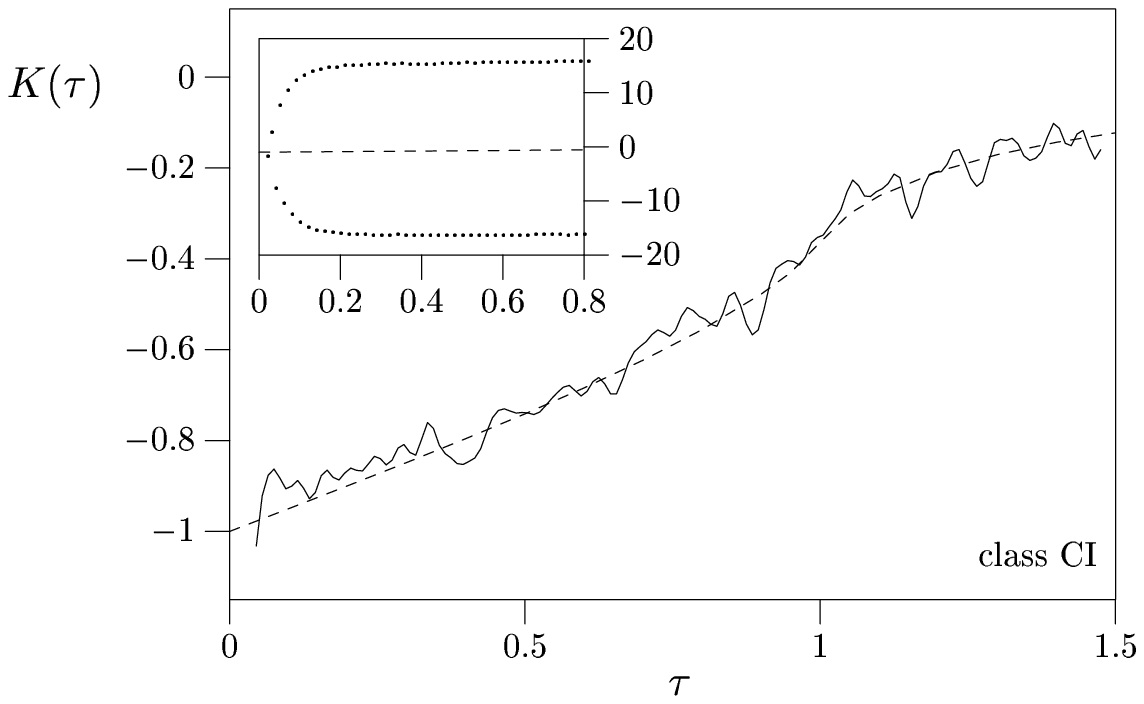}
    \caption{
      Form factors for class $C$ (top) and $C$I (bottom) calculated
      numerically for a star graph (full lines) with $N=100$ bonds
      (averaged over $50\,000$ realizations and a short time interval
      $\Delta t \ll t_H$) 
      and as obtained from 
      the Gaussian random-matrix ensembles $C$-GE and $C$I-GE
      (dashed lines) in dimensionless time $\tau=t/t_H$. 
      The insets give the coefficients $K_m$ as a
      function of ``time'' $m/N$.}
    \label{fig:numerics}
  \end{center}
\end{figure}

Following previous work on quantum graphs \cite{ksmila}, we write the
density of states in $k$ space as 
\begin{equation}
  \rho(k)=\rho_{\mathrm{av}}+\delta \rho(k)
\end{equation}
with 
$\rho_{\mathrm{av}}
= 2 N L/\pi$ and obtain the exact trace formula
\begin{equation}
  \delta\rho(k)=\frac{1}{\pi} \mathrm{Re} \sum_p t_p A_p e^{\ImgI
  S_p+\ImgI\chi}
\end{equation}
as a semiclassical sum over periodic orbits $p$ of
the graph. Here, periodic orbits are defined as a sequence
$i_1,i_2,\dots,i_l$ of peripheral vertices, with cyclic permutations
identified. Since the particle type changes at the peripheral
vertices, the sequences must have even length $l=2m$.  The primitive
traversal ``time'' \cite{lang} of a periodic orbit is $t_p=4mL/r$ (where
$r$ is the repetition number), the stability amplitude is $A_p=1/N^m$,
and the action is 
\begin{equation}
  S_p=4 m k L + \sum_{j=1}^{2m}(-1)^{j+1} 2\pi
  \frac{i_ji_{j+1}}{N}.
\end{equation}
The accumulated Andreev phase is 
\begin{equation}
  \chi =
  -m\pi-\sum_{j=1}^{2m} (-1)^{j+1}\alpha_{i_j}.
\end{equation}
Then, the form factor
becomes 
\begin{equation}
  \begin{split}
    K(t)&=2 \int_{-\infty}^{\infty} dk\, e^{-\ImgI kt} \langle
    \delta \rho (k)\rangle\\
    &= \frac{t_H}{N} \sum_{m=1}^\infty K_m\, \delta(t-\frac{m}{N}
    t_H )
  \end{split}
\end{equation}
with the Heisenberg time $t_H=4LN$ ($\langle \cdot \rangle$
denotes the average over Andreev phases). The coefficients can be
written as a sum over periodic orbits $p_m$ with $2m$ Andreev
reflections:
\begin{equation}
  K_m=2 \sum_{p_m} \frac{m}{r}\left\langle A_p e^{\ImgI S_p(k=0)+\ImgI
      \chi} \right\rangle.
\end{equation}
$K_m$ can be viewed as a form factor in discrete time $m/N$.

\begin{figure}
  \begin{center}
    \includegraphics[width=0.4\linewidth]{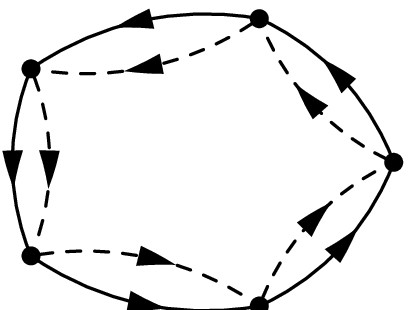}\qquad
    \includegraphics[width=0.4\linewidth]{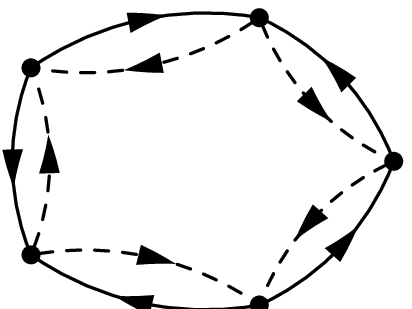}
    \caption{
      Periodic orbits contributing in the self-dual approximation (at
      $m=5$).  The vertices in the diagram correspond to peripheral
      vertices of the original star graph, full and dashed lines
      represent electron and hole propagation.  In class $C$, only the
      left diagram contributes. In class $C$I, the right diagram gives
      $m$ additional contributions as the turning point can be any of
      the $m$ vertices.}
    \label{fig:diagorbits}
  \end{center}
\end{figure}

For graphs in class $C$, only those periodic orbits survive the
average over Andreev phases that visit each peripheral vertex an even
number of times -- half as incoming electron and half as incoming
hole. In the self-dual approximation, only those orbits contribute
whose total phase due to the scattering matrix of the central vertex
vanishes. As the phase factors due to scattering between bonds $i$ and
$j$ for electrons and holes are complex conjugates of one another,
this requires that the periodic orbits contain equal numbers of
scatterings from $i$ to $j$ as electron and hole. This leads to the
orbits sketched in Fig.\ 4: An \emph{odd} number of peripheral
vertices are visited twice, once as an electron and once as a hole.
First, the peripheral vertices are visited once, alternating between
electrons and holes, and subsequently the vertices are visited again
in the same order but with the roles of electrons and holes
interchanged. Thus, these orbits have the same structure as the
self-dual orbits discussed above for the Andreev billiard.  We have
$A_p=1/N^m$, $S_p = 4mkL$, and $\chi=m\pi$. The number of such orbits
of length $2m$ is $N^m/m$, where the denominator $m$ reflects the
identification of cyclic permutations of peripheral vertices. With
these ingredients, we find the short-time result
\begin{equation}
  K_{m,{\mathrm{sd}}}^{C} = -1+(-1)^m 
  \;\; \Rightarrow \;\;
  \overline{K}^{C}_{\mathrm{sd}}(t) = -1,
\end{equation}
where $\overline{K}(t)$ is the time-averaged form factor.  This
reproduces the result predicted by $C$-GE.

For class $C$I, the average over Andreev phases requires only an even
number of visits to each vertex. In the self-dual approximation, this
leads to additional orbits (see Fig.\ 4) and to the result
\begin{equation}
  K_{m,{\mathrm{sd}}}^{C\mathrm{I}} = -1+(-1)^m(2m+1) 
  \;\; \Rightarrow \;\;
  \overline{K}^{C\mathrm{I}}_{\mathrm{sd}}(t) = -1,
\end{equation}
the leading order term for short times predicted by the 
corresponding random-matrix ensemble $C$I-GE.

\section{Andreev billiards without magnetic field}
\label{sec:billiard2}

Finally, we come back to Andreev billiards without magnetic field
(class $C$I).  
As explained above (in section \ref{sec:billiards})
holes necessarily
retrace the electron trajectory, thus leading to non-isolated periodic
orbits and nonuniversal spectral statistics (hard gap). Universal
spectral statistics of $C$I can, however, be found in such Andreev billiards
with $N$ one-channel leads. The reason for this is that Andreev
billiards with $N$ leads containing one channel each can be mapped to
star graphs. The quantization condition for Andreev billiards with $N$
leads is \cite{Richter} 
\begin{equation}
  \det( \mathcal{S}(E) - \mathbbm{1})=0.
\end{equation}
Here
$\mathcal{S}(E)$ is the $N \times N$ Andreev billiard scattering
matrix 
\begin{equation}
  \mathcal{S}(E)=S_{NC}(E)D_- S_{NC}^*(-E) D_+,
\end{equation}
with $S_{NC}(E)$
the scattering matrix describing the coupling of the $N$ channels by
the normal region. The matrices $D_\pm$ describing the Andreev
scattering in the leads are diagonal, $D_\pm=-\ImgI \,
\mathrm{diag}(e^{\mp\ImgI \alpha_i})$, with a specific Andreev phase
$\alpha_i$ for each lead. Time reversal invariance demands
$\alpha_i=0$ or $\alpha_i=\pi$.
Then a detailed correspondence between
billiard and star graph is obtained by substituting $\mathcal{L}S_C
\mathcal{L} \rightarrow S_{NC}(E)$ and $\mathcal{L}S_C^*\mathcal{L}
\rightarrow S_{NC}^*(-E)$ (with a more general central scattering
matrix). Thus, the form factor of these billiards can be obtained in
the self-dual approximation in complete analogy with the star graph.

\section{Conclusions}

We considered the universal spectral sta\-tistics for
ergodic SN hybrid systems belonging to the new symmetry classes, in the
semiclassical approximation. While it was known that semiclassics has
problems in some types of Andreev systems \cite{Melsen,Taras}, we
showed both for billiards and for quantum graphs that the universal
spectral statistics of the random-matrix ensembles $C$-GE and $C$I-GE as
reflected by the appropriately generalized
form factor is correctly reproduced by
semiclassical theory.  An important condition for finding the
universal spectral statistics is that the \emph{combined}
electron-hole dynamics of the Andreev system is classically chaotic.
In particular, this requires that the hole does not retrace the
trajectory of the incoming electron.  In class $C$, this is naturally
the case in magnetic Andreev billiards. We related
the universal features in the density of states
to self-dual periodic orbits which are invariant under electron-hole
exchange.
Our results clarify under
which conditions to expect spectral statistics described by the novel
random-matrix ensembles.

The results presented can be extended 
to the symmetry classes $D$,$D$III and the chiral classes. We
also note that our results for Andreev graphs
remain valid for a rather large class of
central scattering matrices $S_C$. Finally, by going beyond the
self-dual approximation in Andreev graphs, 
it is possible to extract the orbits
contributing to the form factor to linear order in $t$ (weak
localization corrections). These extensions will be discussed
elsewhere \cite{usinwork}.

\acknowledgments

We thank R.\ Klesse, S.\ Nonnenmacher, H.\ Schanz, U.\ Smilansky, and
especially P.\ Brouwer for helpful discussions. This work was
supported in part by SFB 290.


\begin{thebibliography}{}
\bibitem{gauge} Strictly speaking, the phase of the order parameter is
  not a gauge invariant quantity. For class CI, we assume $\mathbf{B}=0$ with
  the gauge $\mathbf{A}=0$ throughout this paper. The discussion of class
  $C$ is valid in any fixed gauge. All physical properties such as the
  form factor are easily shown to be gauge independent.
\bibitem{lang} Although the bond length is a dimensionless quantity,
  we nevertheless refer to it loosely as time.
\end{thebibliography}
\end{document}